# Association of neighborhood disadvantage with cognitive function and cortical disorganization in an unimpaired cohort.

*Apoorva Safai[1], Erin Jonaitis[2,3], Rebecca E Langhough[2,3,4], William R Buckingha[5,6], Sterling C. Johnson[2,3,4], W. Ryan Powell[5,6], Amy J. H. Kind[3,4,5], Barbara B. Bendlin[2,3,4], Pallavi Tiwari[1]*

[1]Departments of Radiology and Biomedical Engineering, University of Wisconsin-Madison, Madison, WI, USA,
[2]Wisconsin Alzheimer's Institute, School of Medicine and Public Health, University of Wisconsin – Madison, Madison, WI, USA
[3]Wisconsin Alzheimer's Disease Research Center, School of Medicine and Public Health, University of Wisconsin – Madison, Madison, WI, USA,
[4]Geriatric Research Education and Clinical Center, William S. Middleton Memorial Veterans Hospital, Madison, Madison, WI, USA,
[5]Health Services and Care Research Program, University of Wisconsin School of Medicine and Public Health, Madison, WI, USA
[6]Department of Medicine, Geriatrics Division, University of Wisconsin School of Medicine and Public Health, Madison, WI, USA

**Abstract:**

**Objective:** Neighborhood disadvantage is associated with worse health and cognitive outcomes. Morphological similarity network (MSN) is a promising approach to elucidate cortical network patterns underlying complex cognitive functions. We hypothesized that MSNs could capture intricate changes in cortical patterns related to neighborhood disadvantage and cognitive function, potentially explaining some of the risk for later life cognitive impairment among individuals who live in disadvantaged contexts.

**Methods:** This cross-sectional study included cognitively unimpaired participants (n=524, age=62.96±8.377, gender (M:F)=181:343, ADI(L:H) =450,74) from the Wisconsin Alzheimer's Disease Research Center or Wisconsin Registry for Alzheimer's Prevention. Neighborhood disadvantage status was obtained using the Area Deprivation Index (ADI). Cognitive performance was assessed through six tests evaluating memory, executive functioning, and the modified pre-clinical Alzheimer's cognitive composite (mPACC). Morphological Similarity Networks (MSN) were constructed for each participant based on the similarity in distribution of cortical thickness of brain regions, followed by computation of local and global network features. We used linear regression to examine ADI associations with cognitive scores and MSN features. The mediating effect of MSN features on the relationship between ADI and cognitive performance was statistically assessed.

**Results:** Neighborhood disadvantage showed negative association with category fluency, implicit learning speed, story recall and mPACC scores, indicating worse cognitive function among those living in more disadvantaged neighborhoods. Local network features of frontal and temporal brain regions differed based on ADI status. Centrality of left lateral orbitofrontal region showed a partial mediating effect between association of neighborhood disadvantage and story recall performance. **Conclusion:** Our findings suggest differences in local cortical organization by neighborhood disadvantage, which also partially mediated the relationship between ADI and cognitive performance, providing a possible network-based mechanism to, in-part, explain the risk for poor cognitive functioning associated with disadvantaged neighborhoods. Future work will examine the exposure to neighborhood disadvantage on structural organization of the brain.

**Keywords:** Social determinant of health, Neighborhood disadvantage, Area deprivation index, Cognitive function, Morphological similarity networks, Alzheimer's disease

**Glossary:** AD = Alzheimer disease; SDOH = Social Determinants Of Health; ADI = Area Deprivation Index; MSN = Morphological Similarity Networks; WADRC = Wisconsin Alzheimer's Disease Research Center; WRAP = Wisconsin Registry for Alzheimer's Prevention; RAVLT-L = Rey Auditory Verbal Learning Test; CF = Category Fluency; WMS-LM = Wechsler's Memory Scale-Logical; TMT-B = Trail-Making Test, part B; MMSE = Mini Mental State Examination; PACC-R= Preclinical Alzheimer's Cognitive Composite–Revised; CAT = Computational Anatomy Toolbox; GM = Gray Matter; WM = White Matter; CSF = Cerebro Spinal Fluid; JSSE = Jenson Shannon Similarity Estimate; PDF = Probability Distribution Function; KLD = Kulback-Leibler Divergence; ROI = region of interest; CC = Clustering Coefficient; DC = Degree Centrality; BC = Betweenness Centrality; LE = Local Efficiency, SW = Small Worldness; GE = Global Efficiency; AST = Assortativity; CPL = Characteristic Path Length; ADRD = Alzheimer disease and related dementia

2. **Introduction:**

Health authorities including the US Department of Health and Human Services and the World Health Organization have emphasized the role of social determinants of health (SDOH), including the neighborhood environment where people are born, live, learn, work, play, worship, and age, as a factor playing a role in cognitive and health outcomes[1,2,3]. The Area Deprivation Index (ADI) is a validated measure of neighborhood disadvantage for the US which includes factors for the theoretical domains of income, education, employment, and housing quality measured within a discrete geographic area[4].

SDOH focused studies have reported associations of neighborhood disadvantage with cognitive decline[5], whole brain and hippocampal atrophy[6,7], cortical thinning[8], as well as Alzheimer's neuropathology[9,10], supporting the notion of neighborhood disadvantage as a risk factor for cognitive decline and neurodegeneration in Alzheimer's Disease (AD). Neuroimaging studies investigating the link between brain patterns, poor cognitive performance and neighborhood disadvantage[3,6,7,8], have been restricted to traditional region level cortical markers. Investigating advanced imaging markers which can elucidate the morphological patterns between brain regions may be especially important, given that the neural mechanisms underlying cognitive functioning are complex and typically involve an interplay of several brain regions and an interregional network pattern. Thus, in the context of examining associations of neighborhood disadvantage with neurocognitive brain patterns, there exists an opportunity to go beyond individual regional measures (e.g. volume, cortical thickness) and explore the interrelated morphological patterns in the brain associated with neighborhood disadvantage.

Recently, morphological similarity networks (MSN) have emerged as a promising alternative to study the cortical organization and inter regional morphological patterns in the brain. MSNs can elucidate the coordinated patterns or similarity in distribution of cortical thickness between brain regions, which are represented in the form of a graph[11]. Network measures computed using graph theoretical techniques can further quantify the local and global topological characteristics or organization of the network, thereby providing network level insights into complex cognitive mechanisms. Our study is based on the hypothesis that MSN may capture the potential effects of neighborhood disadvantage via ADI on inter regional cortical patterns and thereby partially explain the risk of poor cognitive functioning in later life. Our approach leverages the ADI as an area-level measure of neighborhood disadvantage to investigate its association with morphological network patterns and cognitive function. Specifically, we sought to perform the following cross-sectional investigations in a cognitively unimpaired cohort- association of ADI category with (i) cognitive test scores and (ii) MSN measures, and (iii) mediation effect of significantly associated MSN measures in driving cognitive function in neighborhood disadvantaged population.

## 2. Methods:

### 2.1. Study Participants:

The datasets in this study consisted of participants who were enrolled in two large longitudinal studies of AD, namely the Wisconsin Registry for Alzheimer's Prevention (WRAP)[12] study and the Wisconsin Alzheimer's Disease Research Center (WADRC) clinical, which together amount to more than 2500 participants. Both studies determine cognitive status after each visit using NIA-AA criteria for MCI and dementia. Inclusion criteria for these parent studies included (a) fluency in spoken English; (b) adequate visual and auditory acuity to complete study tasks; (c) absence of major psychiatric illness expected to interfere with study participation; and non-demented at cognitive baseline. A subset of cognitively unimpaired participants (n=1529) in both cohorts also completed neuroimaging. Some individuals could not be included in this study due to (missing T1-weighted MRI (n=86); impairment at baseline (n=90); missing cognitive assessment (n=106); difference of more than 6 months between neurocognitive assessment and MRI (n=128), leaving n=1179 eligible for consideration. From that set, we selected only participants in the two top and two bottom deciles of ADI as explained in Section-2.2, which resulted in 537 participants for further analysis. The University of Wisconsin Institutional Review Board approved all study procedures and informed written consent was provided by all participants.

### 2.2. Neighborhood Disadvantage:

Neighborhood disadvantage was measured by the area deprivation index (ADI) which is curated, distributed, and validated at the Census block group level by University of Wisconsin Center for Health Disparities Research via the Neighborhood Atlas[4,13]. ADI is constructed using 17 area-level indicators of poverty, employment, education and housing quality from the 2015 American Community Survey[14]. The ADI scores were determined for individual census block group areas and based on statewide distributions were ranked into relative deciles. Individuals were geocoded to their respective census block group and assigned an ADI score using the most recently reported residential address. The complete list of ADI indicators and detailed method on determination of ADI for WADRC and WRAP cohorts for the current study cohort has been previously described[8]. A higher ADI indicates greater statewide neighborhood disadvantage, whereas lower ADI indicates lower neighborhood disadvantage statewide. Among the ADI scores distributed in deciles from 1-10, we considered individuals from the two lowest (1,2) and two highest (9,10) deciles, based on findings from previous studies which show strongest adverse health effects of neighborhood level factors at the highest level of disadvantage[6,8,9,10].

**2.3. Cognitive Assessment:**

At each visit, participants in the WRAP study complete a comprehensive cognitive battery including six tests in total that evaluated memory, executive function, processing spend, and language ability. Likewise, Wisconsin ADRC participants completed the comprehensive National Alzheimer's Coordinating Center's Uniform Data Set battery[15]. Leveraging tests that overlap in the two cohorts along with published crosswalks mapping same-domain tests to each other, [16] we have made use of several individual cognitive tests and composite scores created from those tests. Individual cognitive test scores for this cross sectional study included the Rey Auditory Verbal Learning Test learning trials 1–5 score (RAVLT-L), Category fluency (CF) test (animal names) score, time to completion on Trail-Making Test, (TMTt, part B), WAIS-R Digit symbol test (WAIS-DS) and Story Memory Delayed Recall (SM-DR) score consisting of a cross-walked score between the Wechsler Memory Scale–Revised Logical Memory delayed recall and Craft Story delayed recall[17]. We also calculated a modified Preclinical Alzheimer's Cognitive Composite (mPACC) using the following three tests: RAVLT-L, TMTt, and LMIIA[16]. This composite was designed to resemble the PACC described by Donohue et al[18]. Composite scores may demonstrate less intraindividual variation in performance compared to individual tests, providing higher sensitivity in detecting exceedingly early cognitive changes related to AD[14,19].

**2.4. MRI Acquisition and Processing:**

High-resolution T1-weighted MRI scans were acquired on 3.0T GE MR750 Scanners using an 8 or 32 channel head coil and a spoiled gradient echo scanning sequence with repetition time = 6.68–8.16ms, echo time = 2.94–3.18ms, inversion time = 400-450ms, flip angle 11–12° and slice thickness of 1×1×1mm. All structural T1w images underwent surface-based analysis using Computation Anatomy Toolbox 12 (CAT12, http://www.neuro.uni-jena.de/cat/) based on Statistical Parametric Mapping 12, which employs a projection-based thickness approach to compute cortical thickness[20]. The complete analysis pipeline for this study is illustrated in Figure 1. Initially, all T1w images were preprocessed as per the standard preprocessing pipeline of CAT12, involving correction of bias field inhomogeneities, segmentation of gray matter (GM), white matter (WM) and cerebrospinal fluid (CSF), and normalization to the MNI template. A fully automated projection approach was implemented for the reconstruction of central surface and computation of cortical thickness, which was measured as the distance between the inner surface (GM/WM boundary) and outer surface (GM/CSF boundary/pial surface) of the GM[21]. Cortical thickness maps were smoothed using a Gaussian kernel with 12-mm full width at half maximum. Images which produced erroneous cortical reconstructions and surface estimates in CAT12 were excluded from further analysis (n=13).

Thus, the total no. of participants containing complete and good quality of imaging, cognitive and ADI data comprised of (n=524, gender- 343 (66%) female, 181 (34%) male; age=62.96 ± 8.37), which were considered for further statistical analysis. Table-1 indicates their detailed demographic information containing sociodemographic characteristics, including age, sex, race/ethnicity, parental dementia history, and educational level of the selected participants.

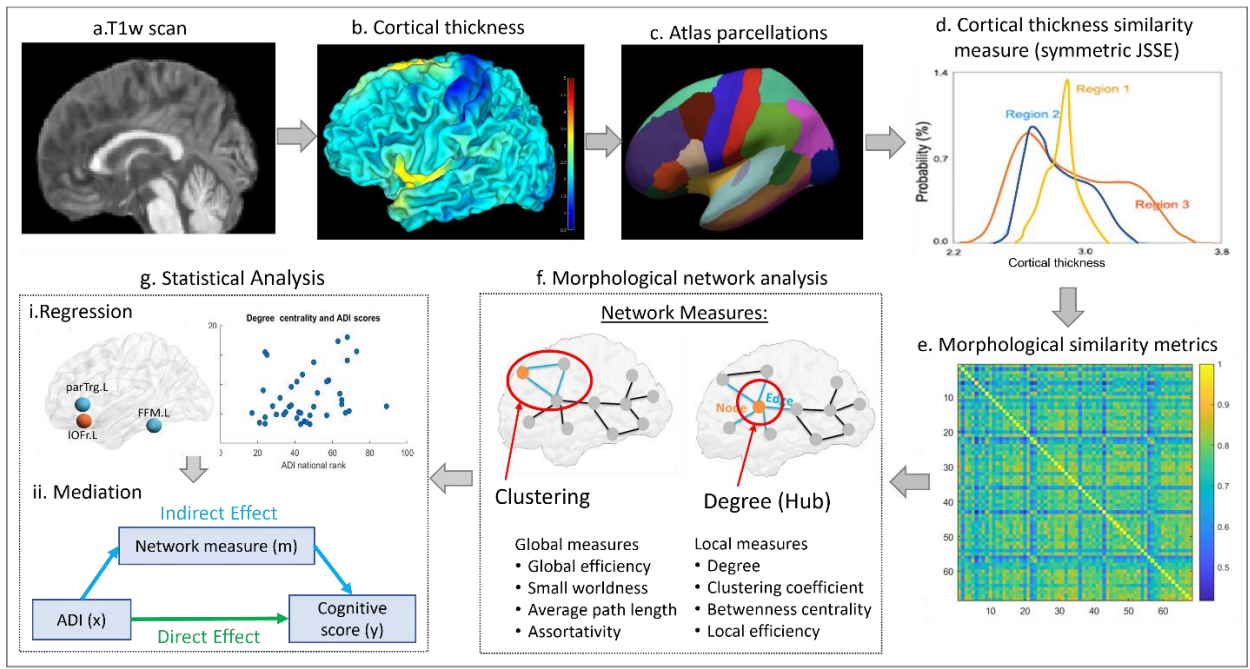

Figure-1: Methodical pipeline framework for investigating the associations between neighborhood disadvantage measured using area deprivation index (ADI), features from morphological similarity networks (MSN) and cognitive performance. Using a) preprocessed T1w scans, c) cortical thickness was computed for 68 cortical brain parcellations from the c) Desikan Killiany atlas d) Jenson Shannon Similarity Estimate (JSSE) was applied between distributions of cortical thickness values of parcellated brain region, in a pairwise manner to construct the e) MSN graph metrics, followed by computation of f) local and global network features using graph theoretical techniques g) Statistical analysis using linear regression was performed to obtain association of

ADI with cognitive scores and MSN features, along with an assessment of the mediating role of MSN features.

**2.5. Construction of Morphological Similarity Networks:**

Networks are graph structures, composed of nodes and edges. We constructed morphological similarity networks (MSN) for each subject individually, where nodes denote brain regions, whiles edges indicate the interregional similarity in cortical thickness. Nodes of the MSN were defined using the parcellations from the Desikan Killiany surface atlas[22], which consisted of 68 left and right hemispheric regions of interest (ROIs) of the cerebral cortex.

Edges indicating morphological similarity between ROIs were estimated using Jensen Shannon Divergence similarity estimate (JSSE)[23]. Firstly, the vertex wise cortical thickness values were extracted for each ROI, followed by a kernel density estimation to obtain the probability density function. Using this resultant probability density function, a probability distribution function (PDF) was obtained for each ROI. Given, two PDF's P and Q for a pair of ROIs, JSSE is computed as the average of Kulback-Leibler Divergence (KLD) between P and M, where M is the average of P and Q. The formulation for KLD based JSSE is given as follows-

$$JS(P\|Q) = 0.5\ (KLD(P\|M) + KLD(Q\|M));\ M(i) = 0.5(P(i) + Q(i))$$

Where, KLD between two probability distributions P and Q is calculated as -

$$KLD(P\|Q) = \sum_i P(i) \log \frac{P(i)}{Q(i)}$$

In this study, JSSE was used in the following manner, as a similarity measure[ref]-

$$JSSE = 1 - \sqrt{JS(P\|Q)}$$

Unlike the asymmetric KLD measure, JSSE is a symmetric metric, with values in the range of 0 to 1, making it a more reliable measure to characterize the morphological similarity between ROIs[24]. Higher JSSE value indicates that the cortical thickness distribution of two ROIs are closer and signifies higher similarity. Thus, the morphological network based on JSSE between 68 brain regions resulted into 68x68 symmetric graph metric.

**2.6. Network Measures:**

Graph theoretical techniques to characterize the local and global topology of the MSN were implemented on symmetric JSSE graph metric using the GRETNA toolbox[25]. To avoid the threshold selection bias, a sparsity-based threshold selection technique was employed, where sparsity was defined as the ratio of the number of actual edges divided by the maximum possible number of edges in a graph. A set of binary adjacency matrices were obtained in the sparsity range from 0.05 to 0.5 at an interval of 0.05. This sparsity range was selected as networks tend to be more fragmented and not fully connected at lower sparsity thresholds, while at higher thresholds they are less random and more likely to maintain small world architecture[26,27]. The local and global network measures were calculated at each sparsity level and were integrated by calculating their area under the curve for further statistical analysis.

For each subject, using graph theoretical measures four local and four global features were computed, which characterized network integration and segregation respectively. Local features such as clustering coefficient (CC), degree centrality (DC), betweenness centrality (BC) and local efficiency (LE) were computed for each of the 68 nodes of the Desikan Killiany atlas. Global features such as small worldness (SW), global efficiency (GE), assortativity (AST) and characteristic path length (CPL) were computed for the entire graph. To determine whether the MSNs were non randomly organized, all global measures were independently normalized by the corresponding mean of 100 randomly generated networks[25].

## 2.7. Statistical Analysis:

Statistical analyses were performed using the R software (version 4.1.3) and the PROCESS macro library in Python[28,29] was used for mediation analysis. The sample characteristics across the low and high ADI cohort were evaluated using a student's t-test as shown in Table 1. Linear regression models were fit to understand the association of neighborhood disadvantage (ADI) separately with cognitive scores and MSN features as illustrated in Figure 1(g). The ADI deciles were coded into two level categorical variables (0 as decile 1,2 and 1 as decile 9,10). For each of the six cognitive outcomes of interest, we examined the linear regression models (at significance level $p<0.05$) as: (i) Cognitive outcome ~ ADI + covariates; where ADI was the independent variable of interest and covariates included age, gender, years of education and practice (i.e., number of prior completions for that outcome). In addition, for each of the global and local network measures, we examined models of network measure (at significance level of $p<0.01$) as: (iia)$MSN_{global(n=4)}$ ~ ADI + covariates, (iib)$MSN_{local(n=4 \times 68)}$ ~ ADI + covariates; where covariates included age, gender and years of education. We also evaluated the association of average cortical thickness of 68 regions with ADI and aforementioned covariates using similar regression model. Mediation analysis was conducted to evaluate a hypothesized causal pathway between neighborhood disadvantage and cognitive function via cortical organization which was ascertained through MSN features. For mediation analysis, we considered MSN features that were statistically significant in regression models (ii), with a higher significance level of p<0.005 considered for $MSN_{local}$ features, owing to the multiple regressions conducted through the model (iib). We obtained the regression parameter estimates for pathways of mediation model and quantified the indirect effect of MSN features on relationship between ADI and cognitive function using product of coefficients method[28]. To assess statistical significance of the indirect mediating effect, a non-parametric bootstrapping (iterations=10,000) approach was implemented, and 95% confidence intervals (CIs) along with beta coefficients, were considered for significance level of p<0.05.

## 3. Results:

### 3.1: Participant demographics:

The detailed demographic characteristics of total study population, and by level of neighborhood disadvantage (Low_ADI and High_ADI) are provided in Table 1. The total study population comprised of 65% females with average years of education as 16.45±2.43. Of the total sample size, 450 participants (85.9%) lived in least disadvantaged (ADI decile=1,2) neighborhoods, whereas 74 participants (14%) lived in a highly disadvantaged (ADI decile=9,10) neighborhood, relative to their state of residence, with the low neighborhood disadvantage group showing significantly higher education level and cognition. The majority of total as well as ADI based sub samples of the study cohort consisted of white individuals, with the high neighborhood disadvantaged sample containing relatively black or African American individuals compared to other sub samples as depicted in Table 1. On average, the time between cognitive examination and MRI scanning for all participants was less than 3 months, and the participants from least and highest disadvantaged neighborhood did not significantly differ on practice effect (number of attempts) during cognitive examination.

| Variables | Total study population | Samples with Low_ADI | Samples with High_ADI |
|---|---|---|---|
| Sample size | 524 | 450(85.9%) | 74(14.1) |
| Female, N (%) | 343(65%) | 287(63.7%) | 56(75.7%)* |
| Age (years: mean±SD) | 62.96±8.37 | 62.97±8.26 | 62.93±9.11 |
| Education (years: mean±SD) | 16.45±2.43 | 16.63±2.38 | 15.33±2.45* |
| APOE e4 positive | 38% | 38% | 38% |
| Parental dementia history | 32% | 29.7% | 46% |
| Primary race (% within each sample) (White / Black or African American / Hispanic /Asian) | 88 / 11 / 1.3 / 0.6 | 91 / 7.5 / 1.5 / 0.6 | 69 / 31 / 0 / 0 |
| Cognitive Score (mPACC) | 0.04±1.12 | 0.11±1.06 | -0.43±1.35* |
| Number of exposures to cognitive tests (practice effect): (median (min-max)) | 0 (0-7) | 0 (0-7) | 1 (0-6) |
| Absolute time between MRI scan and cognitive test(months) | 0.20±0.17 | 0.21±0.17 | 0.16±0.15 |

Table-1: Detailed demographics characteristics of the total population included in this study, as well as of the population living in least (ADI decile-1,2) and highest (ADI decile-9,10) disadvantaged neighborhoods. Difference in characteristics between least and highest neighborhood disadvantaged population was tested using students t-test and chi- square test for gender (* indicating $p<0.005$).

### 3.2. Neighborhood disadvantage and association with cognitive performance:

Neighborhood level disadvantage (ADI) was negatively associated with five of six cognitive outcomes. Participants from highly disadvantaged neighborhood tended to have lower average scores across all cognitive tests. Of the six cognitive test scores, neighborhood disadvantage (ADI) showed significant association with five tests- mPACC ($\beta=-0.47$, $p<0.001$), CF ($\beta=-1.58$, $p=0.023$), TMTt ($\beta=18.47$, $p<0.001$), WAIS-DS ($\beta=-3.68$, $p=0.007$) and SMD ($\beta=-1.66$, $p<0.005$) after controlling for age, sex, education and practice as shown in Table-2. On TMTt and SMD scores, neighborhood disadvantage had a more robust relationship as indicated through the large magnitude of their effect sizes ($\beta$ coefficient) as well as p-value $<0.001$. The distribution of all cognitive scores across the low and high ADI deciles are shown in supplementary material (Figure s1)

| | Intercept | ADI | Age | Gender | Education | Practice |
|---|---|---|---|---|---|---|
| mPACC | 1.26 (0.49) | -0.47 (0.13) | -0.05 (0.005) | -0.49 (0.09) | 0.12 (0.02) | 0.10 (0.02) |
| | p<0.05 | p<0.001 | p<0.001 | p<0.001 | p<0.001 | p<0.001 |
| CF | 26.67 (2.62) | -1.58(0.69) | -0.16 (0.03) | 0.41 (0.51) | 0.37 (0.10) | 0.25 (0.13) |
| | p<0.001 | p<0.05 | p<0.001 | p=0.415 | p<0.001 | p<0.05 |
| WAIS-DS | 86.74 (4.86) | -3.63 (1.35) | -0.62 (0.06) | -2.95 (0.96) | 0.55 (0.20) | 0.47 (0.23) |
| | p<0.001 | p<0.01 | p<0.001 | p<0.005 | p<0.005 | p<0.05 |
| TMTt | 6.98 (16.00) | 18.47 (4.38) | 1.70 (0.19) | 1.46 (3.13) | -2.98 (0.63) | -2.97 (0.78) |
| | p=0.662 | p<0.001 | p<0.001 | p=0.639 | p<0.001 | p<0.001 |
| SMD | 12.36 (1.95) | -1.66 (0.51) | -0.07 (0.02) | -0.79 (0.38) | 0.34 (0.07) | 0.01 (0.09) |
| | p<0.001 | p<0.005 | p<0.005 | p<0.05 | p<0.001 | p=0.912 |
| RAVLT | 61.17 (4.19) | -1.83 (1.09) | -0.40 (0.05) | -6.10 (0.81) | 0.85 (0.16) | 1.20 (0.19) |
| | p<0.001 | p=0.09 | p<0.001 | p<0.001 | p<0.001 | p<0.001 |

Table-2: Results of regression models for association between neighborhood disadvantage (ADI) and cognitive test scores, with age, gender (coded as male-1, female-0), education(years) and practice effect on cognitive test as covariates.

### 3.3. Neighborhood disadvantage and association with cortical MSN features:

The local network features such as centrality, clustering and local efficiency showed significant association with neighborhood disadvantage (ADI) after controlling for age, sex and education, as shown in Table 3. Among the 68 cortical regions, centrality of left hemispheric cortical regions showed a significant (p<0.01, FDR uncorrected) association with ADI, as illustrated in Figure 2. Specifically, centrality of left fusiform (β=-3.74, p=0.009), was negatively related with ADI, while centrality of left lateral orbitofrontal region (β=3.95, p=0.007) showed a positive association with ADI. The average cortical thickness values of these regions, however, did not show any significant (p<0.01) association with ADI (shown in supplementary material Table s3).

| MSN Features | Intercept | ADI | Age | Gender | Education |
|---|---|---|---|---|---|
| Fusiform_L (Ffm.L)_(BC) | 0.13 (5.23) | -3.74 (1.44) | 0.17 (0.05) | 2.86 (1.06) | -0.20 (0.21) |
| | p=0.979 | p<0.01 | p<0.005 | p<0.01 | p=0.336 |
| Lateralorbitofrontal_L (lOFr.L)_(BC) | -2.99 (5.25) | 3.95 (1.44) | 0.12 (0.05) | 0.46 (1.06) | 0.17 (0.21) |
| | p=0.568 | p<0.01 | p<0.05 | p=0.662 | p=0.415 |

Table-3: Results of regression models for association between neighborhood disadvantage (ADI) and local MSN features, with age, gender (coded as male-1, female-0), education (years) as covariates. Results represented as β-coefficients (Standard Error) for each MSN feature.

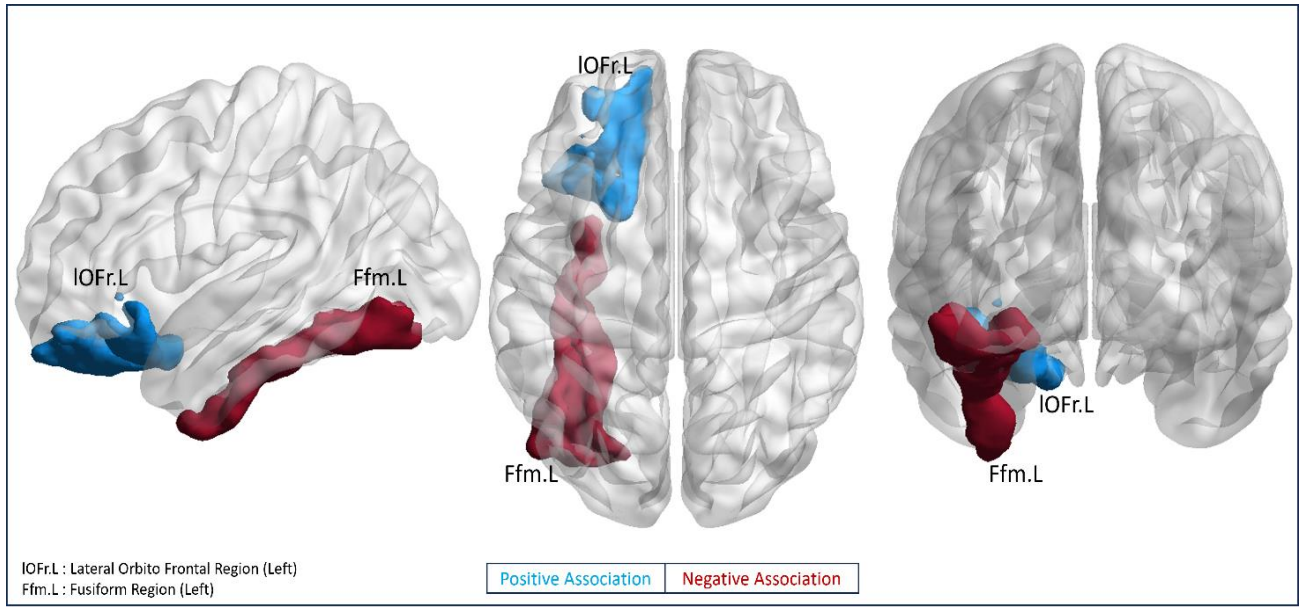

Figure-3: Illustration of cortical brain regions whose local network features showed significant association with neighborhood disadvantage (ADI) after controlling for age, gender, and education (years) as covariates.

of . Global network features did not yield significant association with neighborhood disadvantage status (ADI). Overall, neighborhood disadvantage showed variable association patterns with centrality of frontal and temporal brain regions. The MSN graphs for low and high ADI deciles with respect to their cognitive performances are displayed in Figure-S2 of the supplementary section..

### 3.4. Mediating effect of MSN on neighborhood disadvantage and cognitive function:

We evaluated mediation models, specifically for the local network patterns of frontal and temporal regions that showed an association with neighborhood disadvantage, to assess their involvement in cognitive functions. Results from the mediation analysis indicate that, the negative association of neighborhood disadvantage (ADI) with SM-DR scores was indirectly mediated (-0.176 [-0.463 to -0.023], p = 0.037) through centrality of left lateral orbitofrontal region. Figure 4b represents the complete parameter estimates for the significant mediation effect, indicating that neighborhood disadvantage had a partial mediating effect through increased centrality of left lateral orbitofrontal region on recall mechanism in episodic memory (SM-DR) tasks, which is also found to be impaired in early stages of AD.

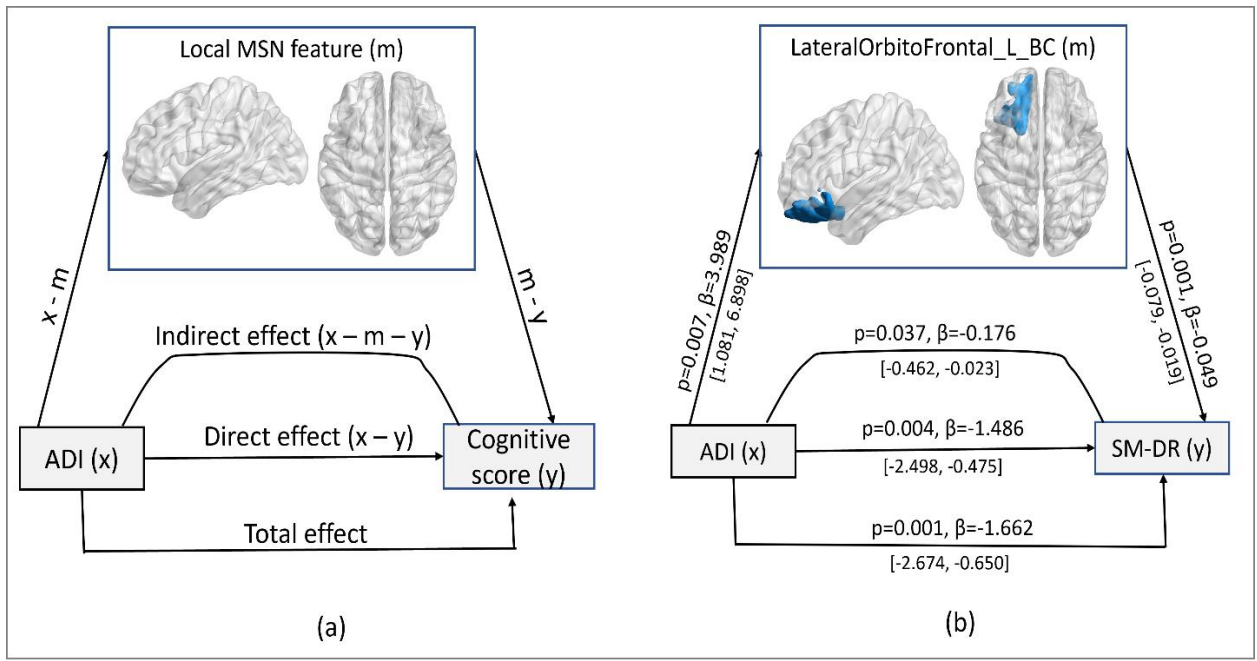

Figure-4: (a) Hypothetical model for mediation on the effect of neighborhood level disadvantage on cognitive function via morphological network patterns. (b) Result of mediation model illustrating the indirect effect of centrality of left lateral orbitofrontal region on the association between ADI and performance on SM-DR scores. 95% confidence interval constructed using nonparametric bootstrapping are indicated as [low, high].

## 4. Discussion:

In this study, we conducted a cross-sectional analysis on neuroimaging data derived from two extensive cohorts from Wisconsin's ADRD research, to investigate the associations of neighborhood disadvantage status which was determined using the ADI deciles (low-0,1; high-9,10), with cognitive performance and morphological similarity networks (MSN). We observed that living in a highly disadvantaged neighborhood was significantly associated with poor cognitive function as well as with altered cortical organization of the brain indicated through the local MSN features. Specifically, the morphological patterns of frontal and temporal region played a prominent role in driving this association, with frontal morphology emerging as a partial mediator in the link between heightened neighborhood disadvantage and compromised episodic memory profiles.

High neighborhood disadvantage showed a significant association with worse scores on TMT-B, SM-DR, CF, WAIS-DS, SM-DR tests and on mPACC composite score, which are also sensitive to cognitive dysfunction in early AD. TMT-B and WAIS-DS tests evaluate executive function, attention and processing speed, which have previously been observed to diminish in the initial stages of both biomarker-defined AD and cerebral small vessel disease[30]. On the other hand, CF and SM-DR tests assess semantic and episodic memory function respectively, which are found to be impaired in early AD and MCI[31,32]. The lower performance in these cognitive domains may signal risk for possible progression to dementia-AD or vascular dementia, but additional study including longitudinal follow-up is needed. The cognitive outcomes observed in this study align most closely with previous research[33-36] examining health outcomes with neighborhood disadvantage which have also shown more robust associations when considering categorical indicators of neighborhood disadvantage.

Existing neuroimaging studies on SDOH have shown neighborhood disadvantage to be associated with brain atrophy[6,7] and longitudinal cortical thinning[8] upon analysis of specific AD related regions. However, they do not report evidence for any significant association with cortical thickness at a cross-sectional level. In our analysis, the left fusiform or lateral orbitofrontal region yielded a significant ($p<0.01$) association with respect to their morphological patterns, but not in terms of their average cortical thickness. While only the average cortical thickness of right entorhinal showed negative association ($\beta=-0.106$, $p<0.01$) with ADI, albeit with a much lower effect size compared to the significant morphological network patterns, as shown in Table-S1 (in supplementary material). This association of average cortical thickness of entorhinal region with ADI also did not yield any significant mediating effect, implying its possible relation to neighborhood disadvantage in mechanism other than cognition. Our investigation of morphological networks was based on the rationale that complex cognitive mechanisms typically involve an interplay of different brain regions, rather than a singular regional dominance, which can be effectively ascertained through network analysis. Moreover, recent studies have also reported cortical disorganization, variably at global as well as local level in mild cognitive impairment and AD, using morphological networks[37,38]. The fusiform region which showed an association with ADI in our analysis, is typically involved in facial recognition[39]. Previous neuroimaging studies have demonstrated early signs of atrophy[37] and cortical thinning[40] of fusiform regions in AD patients, indicating that cortical alterations in these regions may increase the risk of cognitive impairment related to AD. Moreover, neuroimaging studies also suggest the involvement of lateral orbitofrontal region in memory encoding and retrieval of temporal aspects of episodic memory[41]. Thus, the frontal and temporal regions whose cortical patterns were linked with neighborhood disadvantage in this study, also play a role in severe cognitive impairment.

Among the significant morphological patterns that were associated with neighborhood disadvantage, only the centrality of left lateral orbitofrontal region was found to serve as a partial mediator in the relationship between high neighborhood disadvantage and poor memory. These results lend support to a plausible theoretical pathway whereby high neighborhood disadvantage advances cognitive impairment through degeneration in AD-related cortical regions, that is reflected in their abnormal morphological patterns. Specifically in this case, the lateral orbitofrontal region is known to be involved in retrieval mechanism of episodic memory which is assessed through SM-DR test, thereby indicating that the poor memory profiles of individuals in disadvantaged neighborhood may be partially driven by morphological patterns or cortical disorganization of lateral orbitofrontal region. On the other hand, the fusiform region that showed a significant association with neighborhood disadvantage, with no mediating effect on the assessed cognitive tests, could possibly be linked with other cognitive or social abilities. More research is needed in these areas to offer a deeper understanding of these associations.

A few limitations must be acknowledged when interpreting the outcomes of this study. Notably, the study cohort comprised predominantly of white individuals with high levels of education and relatively lower disadvantaged status, residing in the affluent midwestern United States as shown in Table 1. Additional study is needed in more demographically and socioeconomically diverse cohorts. In interpreting the role of neighborhood disadvantage on cognitive functioning, it is essential to consider the temporal nature of extended exposure to neighborhood disadvantage and the onset of neurodegeneration and cognitive decline. Although our analysis was based on a single assessment of neighborhood disadvantage at participant's most recent residential address, ongoing efforts aim to construct residential histories and explore the effects of duration and timing of exposure. It is also important to investigate specific structure inequities that underlie high ADI neighborhoods, including discriminatory zoning policies like redlining, to further understand the root causes of the associations noted herein and to better identify potential interventions towards improved population brain health. Additionally, even though the effect size of associations mentioned in this study are moderately high, the significances for multiple MSN features do not pass FDR correction and hence should be considered as indicative of marginal association trends.

Future studies with a larger and more balanced ADI cohort will involve further validating these findings.

Expanding upon existing literature, this study advances our understanding of the relationship between neighborhood disadvantage, cortical organization, and cognitive function in several ways: 1) presenting preliminary evidence of an association between neighborhood-level disadvantage and cortical networks related to cognitive function, 2) highlighting network level signatures indicating the patterns of morphological alterations in frontal and temporal regions linked with poor cognitive performance in highly disadvantaged neighborhood, 3) utilizing a validated and comprehensive multidimensional construct of neighborhood disadvantage (ADI), rather than a single construct measures, and lastly, 4) the geographic tools employed in this study have been extensively utilized to determine the relative ADI decile of every Census block group across the United States including Puerto Rico and are publicly accessible through the Neighborhood Atlas, facilitating interdisciplinary research in this domain. A potential future scope of the study involving investigating the impact of duration and timing of exposure to neighborhood disadvantage across the life-course, along with longitudinal assessment of ADI with imaging and cognitive markers, could provide stronger evidence for a potential directional association between neighborhood disadvantage, neurodegeneration, and cognitive decline. Moreover, policy initiatives aimed at enhancing community infrastructure may offer valuable opportunities to directly examine causal pathways between neighborhood disadvantage, neurodegeneration, and cognitive decline in middle to older age cohorts.

## 5. Conclusion:

Our cross-sectional study highlights the potential role of SDOH such as neighborhood disadvantage by ADI in later life cognitive impairment as well as cortical disorganization of the brain. The morphological network patterns indicative of cortical disorganization and the poor cognitive performance observed among individuals residing in the most disadvantaged neighborhoods suggest the need for heightened clinical awareness, potentially also including regular screening within this vulnerable population for early signs of MCI or dementia. By delving deeper into the biological pathways of neighborhood disadvantage, and cognitive impairment, clinicians, researchers, and policymakers stand to gain valuable insights for early screening of MCI and targeted strategies for prevention of ADRD.

**Acknowledgements:** The authors thank the researchers and study staff at the WRAP and WADRC for assistance with recruitment, data collection, and data processing; and the WRAP and WADRC study participants.

**Fundings:** This project was supported by National Institute on Minority Health and Health Disparities Award (R01 MD010243 [PI Kind])

**Supplementary Materials:**

| Cortical thickness | Intercept | ADI | Age | Gender | Education |
|---|---|---|---|---|---|
| i. Entorhinal_R (Ent.R) | 3.81 (0.14) | -0.1 (0.04) | -0.005 (0.001) | 0.03 (0.03) | 0.001 (0.006) |
| | $p<0.0001$ | $p<0.01$ | $p<0.005$ | $p=0.243$ | $p=0.832$ |
| ii. Fusiform_L (Ffm.L) | 2.66 (0.05) | -0.007 (0.01) | -0.002 (0.0005) | -0.003 (0.01) | 0.0007 (0.002) |
| | $p<0.0001$ | $p<0.584$ | $p<0.0001$ | $p=0.733$ | $p=0.721$ |
| iii. Lateralorbitofrontal_L (lOFr.L) | 2.81 (0.05) | -0.001 (0.01) | -0.003 (0.0006) | -0.01 (0.01) | 0.004 (0.002) |
| | $p<0.0001$ | $p=0.905$ | $p<0.0001$ | $p=0.147$ | $p<0.05$ |

Table-S1: Results of regression models for association between neighborhood disadvantage (ADI) and average cortical thickness at (i) significance of $p<0.01$ and for regions (ii,iii) which showed significance ($p<0.01$) on local MSN features, with age, gender (coded as male-1, female-0), education(years) as covariates. Results represented as β-coefficients (Standard Error).

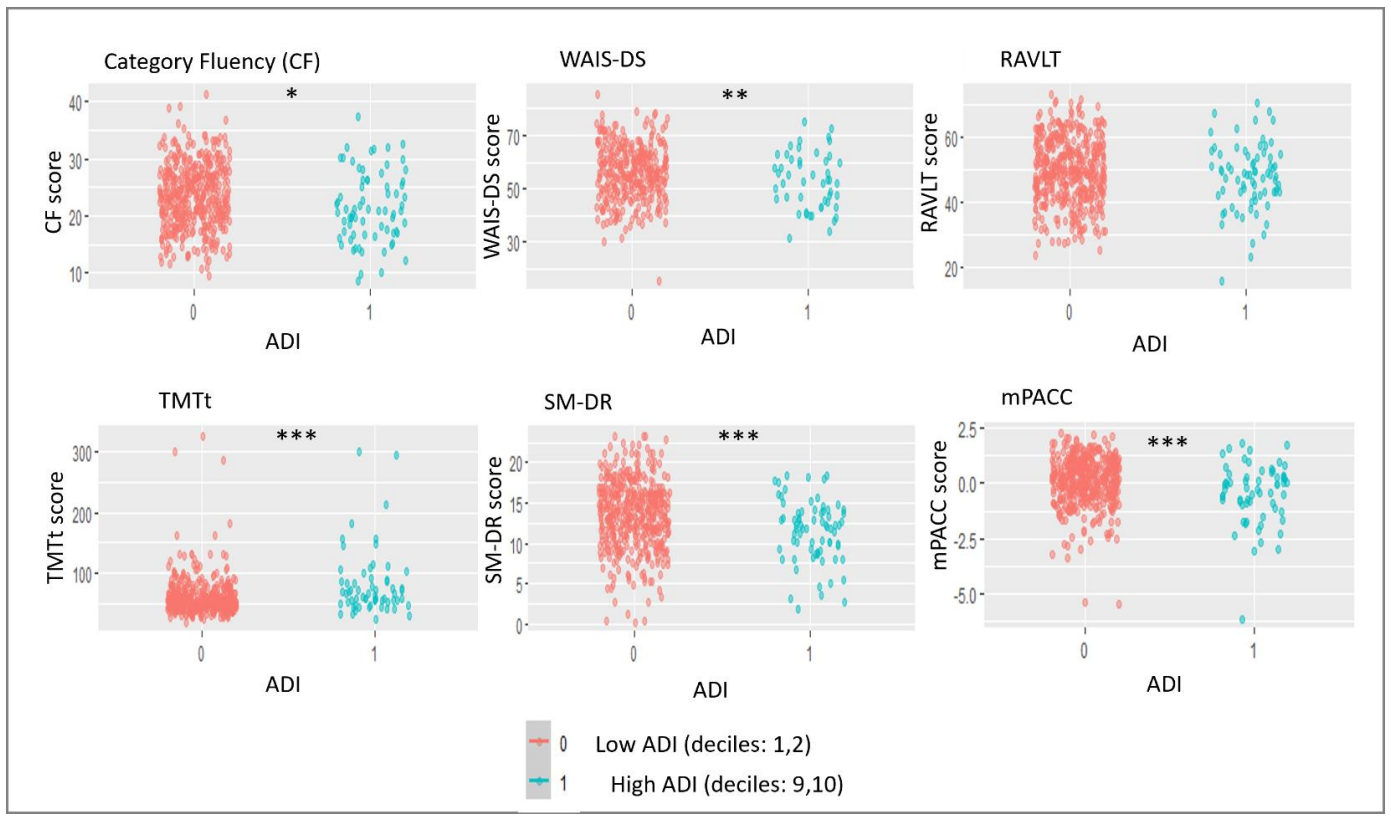

Figure-S1: Plots depicting distribution of cognitive scores across for low (decile-1,2) and high (decile-9,10) ADI populations. Significant associations between ADI and cognitive scores obtained using regression models: Cognitive score (y) ~ ADI (x) + Age+Gender+Education+Practice effect (covariates) are indicated *$p<0.05$, **$p<0.01$, ***$p<0.001$ in the above plot.

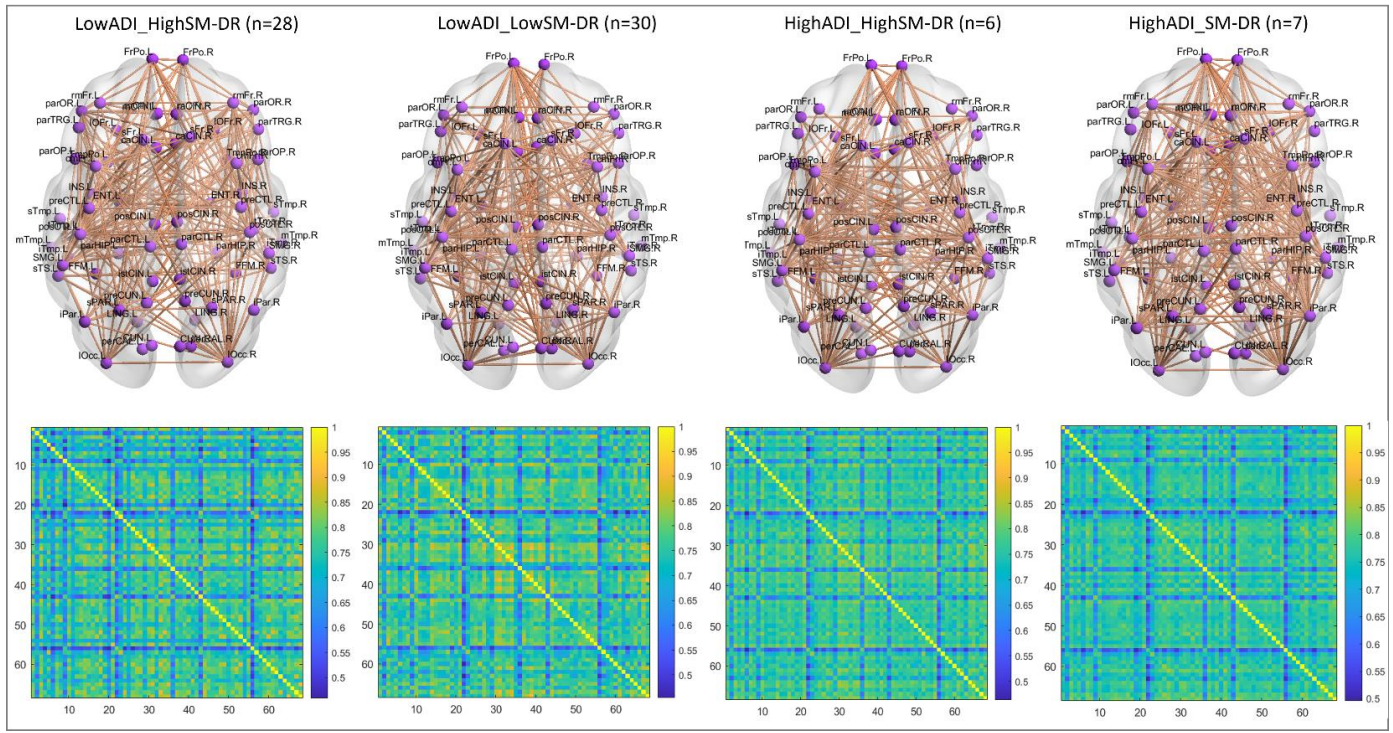

Figure-S2: Visualization of MSN graphs stratified by ADI (low-deciles 0,1; high-deciles 9,10) and cognitive performance on SM-DR tests, with top row indicating top 10% of edges with high value, while bottom row indicates the fully dense graph representing all edge values.